# CHAT ROOM USING HTML, PHP, CSS, JS, AJAX

**Amey Thakur[1], Karan Dhiman[2]**

[1]*Department of Computer Engineering, University of Mumbai, Mumbai, MH, India*
[2]*Department of Computer Engineering, University of Mumbai, Mumbai, MH, India*

---***---

**Abstract -** *Earlier there was no mode of online communication between users. In big or small organizations communication between users posed a challenge. There was a requirement to record these communications and store the data for further evaluation. The idea is to automate the existing Simple Chat Room system and make the users utilize the software so that their valuable information is stored digitally and can be retrieved for further management purposes. There was no online method of communicating with different users. There were many different interfaces available in the market but this method of using windows sockets to communicate between nodes would be fast and reliable. The main objective of our Simple Chat Room project is to create a chat application that helps different users to communicate with each other through a server connected. This is a simple chat program with a server and can have many clients. The server needs to be started first and clients can be connected later. Simple Chat Room provides bidirectional communication between client and server. It enables users to seamlessly communicate with each other. The user can chat using this chat application. If the user at the other end is active then they can start a chat session. The chat is recorded in this application.*

## 1. INTRODUCTION

We propose an application that allows users to create a chat room with a live server and share messages or talk while on the road. Create an instant messaging system that allows users to interact with one another while still being simple enough for a beginning user to utilise. For example, a real-time chat room (online). Teleconferencing, often known as chat, is a technique of bringing people and ideas "together" despite geographical obstacles. Although the technology has been accessible for many years, its adoption is relatively new. This is an example of a chat server in this small project. The chat application is straightforward. It doesn't require a login, has AJAX-style functionality, and will support numerous users. It consists of two applications: the client app, which runs on the user's device, and the server programme, which is hosted to operate the chat room live over the network. To begin talking, clients should connect to a server where they can practise two types of chatting, public (message is broadcasted to all connected users) and private (between any two users alone), and security precautions were implemented during the latter.

## 2. PROPOSED SYSTEM

### 2.1 Architecture of Chat Room

Application for Chat Rooms A data flow diagram is usually used as a first stage to develop an overview of the Chat Application without going into depth, which could then be worked upon later. Some of the user's flow and related entities described are Chat, Chat History, Smiley Chat, Chat User, Chat Group, Chat Profile, and Delete Conversation. The diagrams below are used to visualize data processing and a better design of the Chat Application process and activity.

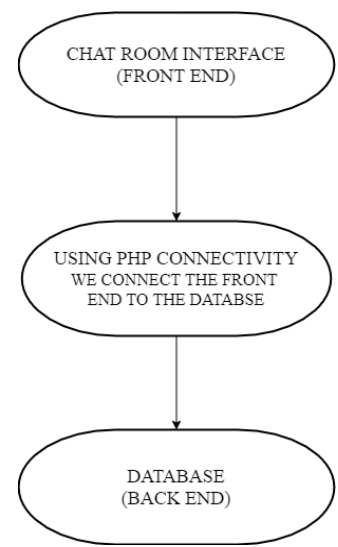

Fig. 1: Flow Diagram

## 2.2 Zero Level Data Flow Diagram (0 Level DFD) of Chat Room Application

This is the Zero Level DFD of an Online Chat Application, in which we have explained the high-level methodology of the Chat Application. It's meant to give you a brief explanation of Chat User, Chat Group, and Delete Chat by representing the system as a single high-level process with relations to various entities such as Chat, Chat History and Chat Profile. It is a high-level overview of the entire Online Chat Application or process that is being studied or modelled.

It should be understood to a number of different groups, including Chat Profile & Chat User. In the zero levels DFD of Chat Rooms Application, we described the high-level flow of the Chat Application system.

High-Level Entities and process flow of Online Chat Application are given in the diagram below:

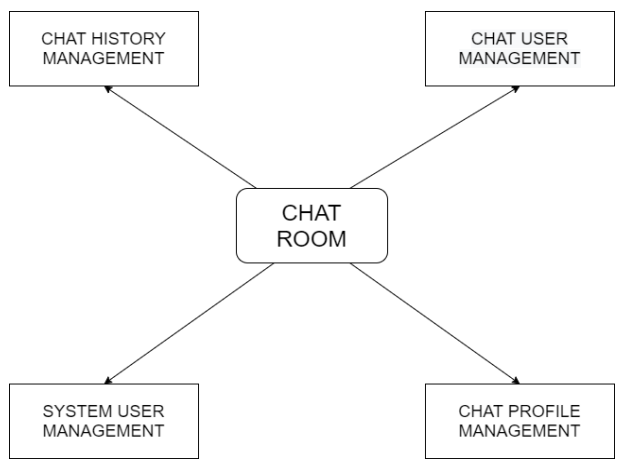

Fig. 2: 0 level DFD

## 2.3 One Level Data Flow Diagram (1 Level DFD) of Chat Room Application

The first level DFD (1st Level) of the Chat Rooms Application provided in this section is segmented into subdivisions (processes), each of which handles one or more data flows to or from an external agent, and which together offers all of the Chat Rooms System application's functions. It also identifies internal data storage like Chat Users and Smiley Chat. DFD Level one is a more in-depth variation of DFD Level two. You'll outline the Chat Application's most key aspects in the diagram given below:

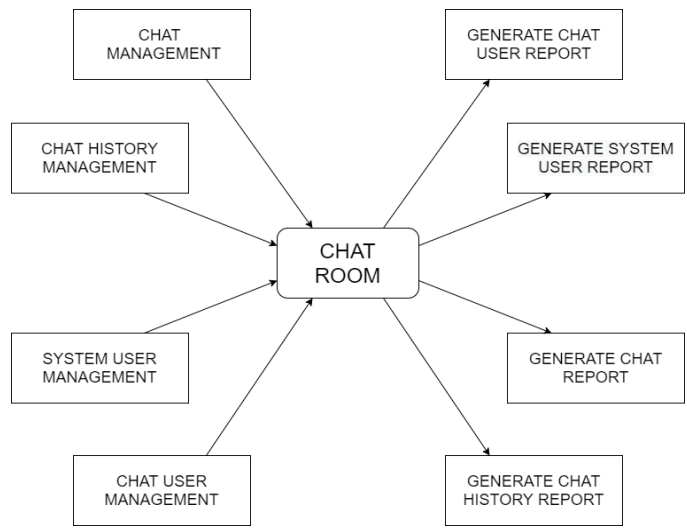

Fig. 3: 1 level DFD

## 2.4 Two-Level Data Flow Diagram (2 Level DFD) of Chat Room Application

DFD Level two then delves further into aspects of Level one of the Chat Application. More Chat Application features may be required to achieve the required amount of detail regarding the Chat Application's operation. The first level DFD (1st Level) of the Online Chat System demonstrates how the system is split into subsystems (processes). More about second-level DFD, one may have more information on Chat Group, Chat User, Smiley Chat, Delete Chat, Chat Profile, Chat History, and Chat.

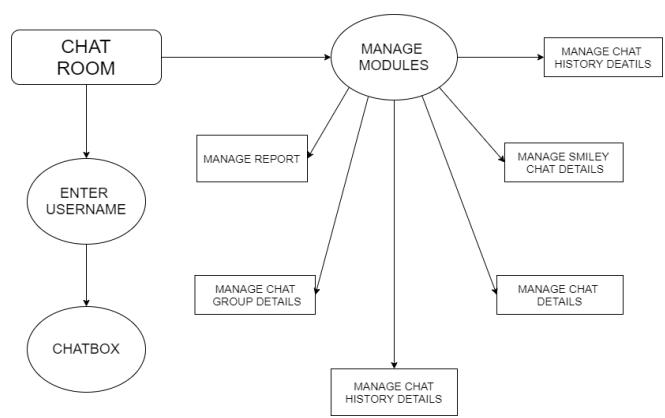

Fig. 4: 1 level DFD

## 2.5 Applications of Chat Room System

Here are some examples of chat room system applications:

1. Google Chat
2. Chat Room on Google Meet
3. Hangouts on Google
4. Microsoft Teams Chat Room
5. Chat Room for Zoom Meeting

## 3. LITERATURE SURVEY

### 3.1 Brief History of Off-Line Computerized System or existing system similar to your chosen system

Between ancient times till the 15th century, communication began to change. While it is exciting, we do have other things to do. So this is what I'll say in conclusion: Prior to the 15th century, communication depended mostly on verbal and minor written communication. People interacted with one another and made handwritten notes for people. Everything improved after the printing. The press was invented to reduce the necessity for all paper messages to be written by hand.

The public postal declaration was made in the 18th century. Before this time period, letters were distributed, but the procedure was broken and inefficient. It could take months before mail to reach, and even when it did, it was deposited at a random public spot.

As the 19th century began, great thinkers focusing on improving communication made great progress. The name telecommunications was developed after electricity was introduced into communication routes. Telecommunications provided people with a way to make long-distance written and verbal communication personal. For all of those who wished to connect orally, the distance became less of an obstacle, permitting greater information to flow.

In the early 1900s, radio and television became valuable sources of information. They not only offered musical delight, but they also alerted listeners with news, sports, and the weather. We all know that email and PCs were just the beginning of the most important innovations in communication. But these origins can't be forgotten. Every communication tool used today was somehow influenced by simple ones from the long and never-ending history of communication.

21st-century communication. When thinking of modern-day communication, keeping the idea of the Information Age in mind is important. The Information Age refers to the transition from industry to information technology. Simply put, everything is digital and knowledge has never been more powerful. People want information, and the best way to get it is through communication tools.

### 3.2 Comparison of Chat Room System with the Previous system

| Sr. No. | Basis | Chat Room | Google chat | Google Meet Chat Room | Microsoft Teams Chat Room | Zoom meeting chat room |
|---|---|---|---|---|---|---|
| 1 | Number of participants | Limitless | 150 | 250 | 250 | 100 |
| 2 | Number of characters | 255 | 160 | 4000 | 1000 | 128 |
| 3 | Cost | Free | Paid | Free Trial | Free Trial | Paid |
| 4 | Anonymity | Yes | No | No | No | No |
| 5 | Launch Year | 2020 | 2017 | 2020 | 2017 | 2013 |

Fig. 5: Comparison of Chat Room System with the Previous system

## 4. PROBLEM STATEMENT

### 4.1 Problem Statement

Develop an application that facilitates the creation of a chat room with a live server for the users to enable sharing messages or chat on the go. Develop an instant messaging system that allows users to connect effortlessly with each other and still being simple to use. I.e. Live chat room on the fly (online).

### 4.2 Problem Solution

Chat Room as a service is a model of communication deployment where the server hosts a live chat room as a service for users on the Internet. Users are admissible to enter the chat room and share messages or interact with each other. The project has been created keeping in mind the fact that the anonymity of the users will not be compromised under any circumstances. Our aim is to identify a solution to use networks to serve people and ideas throughout geographical boundaries.

### 4.3 Chat Room scope and features

Chat Server Application will be a text communication programme that can communicate between two computers via a point-to-point connection. This project's main characteristic is its anonymity. Our project's drawback is that it does not support audio chats. Companies want communication software that allows them to communicate immediately inside their organisation. Because software operates on an intranet within the company, it is extremely safe from outside threats.

## 5. DESIGN AND CONNECTIVITY

### 5.1 Design Process

User experience or (UI) design is the process through which designers create interfaces in software or electronic devices with an emphasis on aesthetics or style. Designers strive to develop interfaces that are both easy to use and enjoyable for users. Graphical user interfaces and various kinds of user interface design are examples of UI design. To design System Layout Architecture we need to have UI and Database and to connect these two using PHP connectivity.

### 5.2 Database Design

A database can be comparable to a sophisticated digital format cabinet. It's what might help us organise all or most of the data throughout our app. We have total control over inserting, altering, and deleting from our database.

Tables, that may be considered as file directories, make databases. Tables are rows of information that may be thought of as independent papers within a file folder. We will add new entries to that database every time an amount is charged, just as we could add further pieces of paper to our file folder.

Our database contains an amount of data, such as the amount, bill number, and date paid. We will be able to collect, preserve, modify, and examine information in our web app due to the database. We do have the capability to obtain information that has been submitted into our database

A query is a command with which we could send to our database that directs it to perform specific operations.

### 5.3 Database connection

```
<?php
    $host = "localhost";
    $user = "root";
    $pass = "";
    $db_name = "chat_info";
    $con = new mysqli($host,$user,$pass,$db_name);

    function formatDate($date)
    {
        return date('g:i a',strtotime($date));
    }
?>
```

## CONCLUSIONS

Chat Room achieves its goal by delivering an exceptionally rich conversation experience. We attempted to keep the UI clear and clean, with no obnoxious or unnecessary embellishments. Design flexibility encourages users to utilise their creativity, and as a result, even inexperienced users may create effective websites. We have used PHP, MySQL, JavaScript, and Ajax to build a dynamic internet messaging system.

There is always an opportunity for improvement in every product, and we attempted to adjust the design accordingly, while still keeping our constraints in mind. During the course of developing this application, we faced a slew of issues and learned how to solve them through study. With the end product, we think our idea was evident and well-presented.